# Crystal Structure and Elementary Properties of $Na_xCoO_2$ (x = 0.32, 0.5, 0.6, 0.75, and 0.92) in the Three-Layer $NaCoO_2$ Family

L. Viciu,[1*] J.W.G. Bos[1], H.W. Zandbergen,[2] Q. Huang,[3] M. L. Foo[1], S. Ishiwata,[1,4] A. P. Ramirez,[5] M. Lee,[6] N.P. Ong[6] and R.J. Cava[1]

[1] Department of Chemistry, Princeton University, Princeton NJ 08540

[2] National Centre for HREM, Department of Nanoscience, Delft Institute of Technology, Al Delft, The Netherlands

[3] NIST Center for Neutron Research, NIST, Gaithersburg, MD 20899

[4]Department of Applied Physics, Waseda Univ., Ookubo, Shinjuku, Tokyo 169-8555, Japan

[5]Bell Laboratories, Lucent Technologies, Murray Hill NJ 07974

[6]Department of Physics, Princeton University, Princeton, NJ 08540

## Abstract

The crystal structures of the $Na_xCoO_2$ phases based on three-layer $NaCoO_2$, with x=0.32, x=0.51, x=0.60, x=0.75 and x=0.92, determined by powder neutron diffraction, are reported. The structures have triangular $CoO_2$ layers interleaved by sodium ions, and evolve with variation in Na content in a more complex way than has been observed in the two-layer $Na_xCoO_2$ system. The highest and lowest Na containing phases studied (x=0.92 and x=0.32) are trigonal, with three $CoO_2$ layers per cell and octahedral Na ion coordination. The intermediate compositions have monoclinic structures. The x=0.75 compound has one $CoO_2$ layer per cell, with Na in octahedral coordination and an incommensurate superlattice. The x=0.6 and x=0.5 phases are also single-layer, but the Na is found in trigonal prismatic coordination. The magnetic behavior of the phases is similar to that observed in the two-layer system. Both the susceptibility and the electronic contribution to the specific heat are largest for x=0.6.

**Introduction**

$Na_xCoO_2$ has been widely studied as solid-state cathode in Na batteries.[1] The discovery of superconductivity ($T_C$=4.5K) in two-layer $Na_{0.3}CoO_2$ intercalated with water in 2003,[2] made the study of the $Na_xCoO_2$ system an active area of research. The degree of filling of the Na layer controls the charge in the $CoO_2$ planes, giving rise to different properties as a function of x. The two-layer form of $Na_xCoO_2$ has been of significant recent interest, as it displays a variety of interesting properties. In addition to the superconductivity in the hydrated phase, a large thermopower (100μV/K at 300K) has been found for x~0.7[3] and attributed to the spin entropy carried by strongly correlated electrons hopping on a triangular lattice.[4] A transition to an insulating state takes place in $Na_{0.5}CoO_2$ at low temperatures that has been attributed to charge ordering.[5]

Four different phases have been previously reported in the thermodynamic $Na_xCoO_2$ chemical system.[6] In all the phases, sheets of edge-sharing $CoO_6$ octahedra are interleaved by sodium ions. The stacking sequence of the oxygen layers gives the number of sheets within a unit cell. Either two or three $CoO_2$ sheets per unit cell are found. Three of the four phases are reported to be three-layer structures, delineated as (1) the α- phase for $0.9 \leq x \leq 1$; (2) the α'-phase for x=0.75 and (3) the β-phase for $0.55 \leq x \leq 0.6$. Only one thermodynamic phase has a two-layer structure, known as the γ phase, for x ~0.7. The coordination of sodium ions in these structures is either octahedral or trigonal prismatic.

These four thermodynamically stable phases in the $Na_xCoO_2$ system can be obtained by classic solid state reactions. Topochemical methods can be used to tune the sodium composition within these structures. Thus, two layer $Na_{0.5}CoO_2$ and $Na_{0.3}CoO_2$ have been obtained by chemical deintercalation of the higher *x* counterparts.[7] In addition, chemical intercalation can be used to increase the sodium content of the γ phase from x=0.7 up to x=1.[7] The crystal structure of two-layer $Na_xCoO_2$ with 0.3≤x≤1 has been extensively studied by Rietveld refinement using neutron diffraction data.[7] In the three-layer structure, however, crystallographic studies are reported only for x=1 (single crystals),[8] and for x=0.67 (polycrystalline powder).[9] At x=1 the reported structure is trigonal (*R-3m* with a=2.889(2) and c=15.609(3)Å) while at x=0.67 the crystal structure is single-layer monoclinic (*C2/m,* with a=4.9023(4), b=2.8280(2), c=5.7198(6) and β=105.964°). Early on, the powder patterns for x=0.5 and x=0.6 were indexed with a monoclinic cell but the structure was not determined.[10]

Here we report a structural study, by neutron powder diffraction analysis, of the $Na_xCoO_2$ phases derived from three-layer $NaCoO_2$, for x=0.92, 0.75, 0.60, 0.51, and 0.32. It is found that the crystal structure changes from one sodium composition to another in an unexpected way. For example, $Na_{0.92}CoO_2$ is trigonal, with Na in octahedral coordination. Deintercalation of this compound using $Br_2$ results in the formation of $Na_{0.3}CoO_2$, which has the same crystal structure. Deintercalation of trigonal $Na_{0.92}CoO_2$ with $I_2$ forms $Na_{0.5}CoO_2$, which has a monoclinically distorted single-layer structure. $Na_{0.6}CoO_2$ has also a single-layer unit cell and is found to be isostructural with the previously reported $Na_{0.67}CoO_2$.[9] $Na_{0.75}CoO_2$ has a complex crystal structure. An average structure for this phase is reported based on the main reflections in the neutron diffraction data, indexed with a monoclinic cell. Susceptibility and heat capacity measurements are also reported for these phases. Although different crystal structures are found for the three-layer as opposed to the two-layer structures, the basic electronic properties of these materials are similar, supporting the general understanding that the electronic systems, dominated by the in-plane character of the $CoO_2$ layers, are highly two dimensional in character.

**Experimental**

Samples of $Na_xCoO_2$ with x=0.92, 0.75, and x=0.60, were obtained as previously described.[6] Stoichiometric amounts of $Na_2O_2$ (Alfa, 93% min) and $Co_3O_4$ (Alfa, 99.7%) were mixed together in an argon filled glove box. The powders were then quickly removed from the glove box and placed in a tube furnace to prevent the hydration of sodium peroxide by air exposure. The temperature was slowly (5°C/min) increased to 550°C, held constant for 16 hours and then slowly (5°C/min) cooled to room temperature under flowing oxygen.

$Na_{0.5}CoO_2$ and $Na_{0.3}CoO_2$ were synthesized by chemical deintercalation of $Na_{0.92}CoO_2$. $Na_{0.5}CoO_2$ was prepared by mixing $NaCoO_2$ with excess $I_2$ (10x) dissolved in acetonitrile. After 5 days of stirring, the product was washed with acetonitrile, dried and stored under argon. Single crystals of $Na_{0.5}CoO_2$ were obtained by chemical deintercalation of $NaCoO_2$ single crystals stirred with $H_2O_2$ at room temperature for 5days. $Na_{0.3}CoO_2$ was obtained by mixing $Na_{0.92}CoO_2$ with a molar excess of 40x bromine dissolved in acetonitrile. The reaction time was 5 days, after which the product was washed with acetonitrile and stored under argon. Minimum exposure to atmospheric conditions is required to prevent water intercalation.

All samples were analyzed by powder X-ray diffraction using Cu K$\alpha$ radiation and a diffracted beam monochromator. Neutron diffraction data were collected on each sample at the NIST Center for Neutron Research on the high resolution powder neutron diffractometer with monochromatic neutrons of wavelength 1.5403 Å produced by a Cu(311) monochromator. Collimators with horizontal divergences of 15', 20' and 7' of arc were used before and after the monochromator and after the sample, respectively. Data were collected in the 2-theta range of 3° to 168° with a step size of 0.05°. The structural parameters were refined using the program GSAS.[11] The neutron scattering amplitudes used in the refinements were 0.363, 0.253, and 0.581 ($\times 10^{-12}$ cm) for Na, Co, and O, respectively. All sodium contents for the phases were determined by the structure refinements, and were in good agreement with those expected from nominal compositions.

Electron microscopy analysis was performed with Philips CM300UT electron microscopes having a field emission gun and operated at 300kV. Electron transparent areas of specimens were obtained by crushing them slightly under ethanol to form a suspension and then dripping a droplet of this suspension on a carbon-coated holey film on a Cu grid or Au grid.

The magnetic susceptibilities were measured with a Quantum Design MPMS SQUID system. Zero field cooled (ZFC) magnetic data were taken between 2K and 300K in an applied field of 5kOe. The specific heat samples were prepared by cold-sintering the sample powder with Ag powder. Measurements were made in a commercial cryostat using the relaxation method. A four-probe method using a Quantum Design PPMS system was used to measure the resistivity of x=0.5 phase in the 5-300K temperature range.[12]

**Results**

The purity of the $Na_xCoO_2$ compounds was confirmed by X-ray diffraction analysis, while structural investigations were performed by neutron diffraction analysis. Only $Na_{0.92}CoO_2$ and $Na_{0.32}CoO_2$ had powder diffraction patterns that could be indexed with a hexagonal cell. Their neutron powder patterns were indexed within the trigonal space group *R-3m* (No. 166), with the cell parameters given in table 1. None of the intermediate sodium compositions maintained the hexagonal structure. The neutron diffraction patterns of $Na_{0.5}CoO_2$ and $Na_{0.6}CoO_2$ were indexed based on centered monoclinic cells in the space group *C 2/m* (No. 12). The cell constants are presented in table 1. The monoclinic cells arise through shifts of the $CoO_2$ planes of approximately an angstrom relative to each other to accommodate changes in the local Na-O coordination. The shifts are in one direction, parallel to the crystallographic *a* axis. The X-ray diffraction pattern of $Na_{0.75}CoO_2$ could be well indexed based on a monoclinic cell (space group *C 2/m*), but the neutron diffraction pattern showed the presence of relatively strong incommensurate superstructure reflections also seen in electron diffraction (described below). The main reflections of the neutron diffraction pattern fit the cell parameters presented in table 1. In the following, the nomenclature introduced earlier to describe layered structures of this type[13] is employed to most easily distinguish the phases: O and P designations refer to octahedral or prismatic coordination, respectively, of Na in the phase, and the numerical designation 1, 2, or 3 refers to the number of $CoO_2$ layers in a unit cell repeat. In this system of nomenclature, the two layer phases commonly studied are P2 compounds, whereas the parent compound for the current studies, $NaCoO_2$, has the O3 type.

**Structural Characterization.**

**x=0.92 and x=0.32. [O3 structure type]** The $Na_{0.92}CoO_2$ and $Na_{0.3}CoO_2$ phases have been found to be isostructural. Their structural analysis by the Rietveld method was carried out in the space group *R-3m*. The sodium ions are on the (0, 0, ½) site, and are coordinated octahedrally to the oxygens from the $CoO_2$ layers. The refined structural parameters for both phases are presented in tables 2 and 3. Sodium contents of 0.921(7) and 0.32(1) were determined by refinement for these two phases. As an example, the structural model for $Na_{0.92}CoO_2$ is shown in figure 1.

**x=0.6 and x=0.51. [P1 structure type]** $Na_{0.6}CoO_2$ and $Na_{0.5}CoO_2$ have similar crystal structures and are found to be isostructural with $Na_{0.67}CoO_2$.[9] This is a single-layer structure (space group *C 2/m*) where the sodium ions are in trigonal prismatic coordination. The structural parameters for $Na_{0.6}CoO_2$ and $Na_{0.5}CoO_2$ are presented in table 4 and table 5 respectively. Whereas in $Na_{0.5}CoO_2$ the sodium ions are found on the *4i* site (x,0,z), in $Na_{0.6}CoO_2$ the sodium ions are displaced from the *4i* site to a more general position *8j* (x, y, z). No such displacement of the sodium ions was detected for $Na_{0.5}CoO_2$ within the standard deviations of the positional parameters. In these structures, each sodium layer of $Na_{0.5}CoO_2$ has only ~25% of a honeycomb-geometry sodium lattice occupied, while in $Na_{0.6}CoO_2$ the sodium ions form a distorted honeycomb lattice only 30% filled. No information about any possible ordering of the Na ions within the partially occupied sites was obtained in the present study. A very good agreement of the sodium content with the nominal compositions is found for both phases by Rietveld refinement: x=0.596(3) and x=0.512(4). As an example, the observed, calculated and the difference plots of the refinement for the x=0.51 phase are presented in figure 2. The idealized crystal structure is shown in figure 1.

**x=0.75. [O1 structure type]** A different crystal structure from the ones described above is found for $Na_{0.75}CoO_2$. The Rietveld refinement was performed in the C *2/m* space group based on a single-layer structure. The crystal structure of the x=0.92 phase was used as a starting model with the atom coordinates in the three-layer structure transformed into the single-layer monoclinic cell. This model gave a very good fit to both the powder X-ray diffraction data ($\chi^2$=1.26, wRp=7.86%, Rp=6.22%) and the main reflections in the neutron diffraction pattern. The refined structural parameters based on the neutron diffraction data are presented in table 6. Figure 3 shows the observed, calculated and difference plots for the refinements based on this model for both X-ray data and neutron data (inset in fig 3). The structural model is presented in figure 1.

Electron diffraction studies were performed to characterize the superstructure observed in the neutron diffraction pattern of $Na_{0.75}CoO_2$. Figure 4 shows an [010] electron diffraction pattern of $Na_{0.75}CoO_2$ taken in nano-diffraction mode with a spot size of about 10 nm. A quite dominant superstructure can be distinguished. The superstructure reflections can be indexed best using four dimensions, resulting in four Miller indices, *hklm*, in which *m* indicates the order of the satellite measured from the nearby main reflection. The *hklm* indexing of two superreflections is shown in figure 4. The superreflections in the electron diffraction patterns are much more visible than in the neutron powder diffraction pattern (they are not seen in the X-ray powder pattern). This is due to the strong dynamic scattering in electron diffraction, which results in an enhancement of the weak reflections compared to the strong ones. The observed superstructure for x=0.75 is relatively strong, since quite a number of 1$^{st}$ order superreflections can be seen in the neutron diffraction pattern (see figure 3). The indexing of the superreflections in the neutron diffraction pattern is based on the 0.330 a* - 0.247 c* modulation vector determined from electron diffraction data. Determination of the incommensurately modulated structure of $Na_{0.75}CoO_2$ from the powder diffraction data is beyond the scope of the present study.

**Magnetic Characterization.** The variation of molar susceptibility with x is presented in figure 5. At temperatures higher than 50K, the temperature dependence of the susceptibility ($\chi$ vs. T) for x≥0.6 follows the Curie-Weiss law, $\chi=\chi_0+(C/T-\theta)$, with the Curie constant (C), Weiss constant ($\theta$) and temperature independent term ($\chi_0$) presented in table 7. The negative Weiss constants indicate antiferromagnetically interacting spins. Given a simple localized picture where only $Co^{4+}$ ions carry the spin moment S=1/2, the effective moments for the 0.92, 0.75 and 0.6 compositions are expected to be 0.49, 0.87 and 1.09 $\mu_B$/(formula unit), respectively. The observed moments are 0.32, 0.88 and 0.82 $\mu_B$/(formula unit) respectively. Although the differences between the calculated values assuming a simple model and the observed ones are not large, developing experiments in the two-layer phase indicate that a more complex electronic and magnetic system is at play (see e.g. ref. 14). It can be inferred, however, that low spin configurations for $Co^{3+}$ and $Co^{4+}$ ions are found in these phases. Small deviations from the Curie law are noted for x=0.92 and x=0.6 below 30K, and a magnetic transition at ~30K is seen for x=0.75, at a similar temperature to that observed for the two-layer x=0.75 phase.[14] For x=0.5, $\chi$ vs.

T has a different shape and no Curie Weiss behavior is found. Two cusps are observed in the $\chi$ vs T data: one at ~88K and another one at ~52K. These two transitions are observed at nearly the same temperatures where transitions are observed for two-layer $Na_{0.5}CoO_2$.[5] In the two-layer structure the two anomalies signal the onset of an insulating state that has been attributed to a charge ordered phase. For x=0.3 in the three-layer structure, $\chi$ is independent of temperature above 75K. A Curie contribution is seen at lower temperatures. The origin of this behavior is not known.

**Resistivity Measurements.** Resistivity measurements were performed on x=0.5 single crystals (figure 6). A transition to insulating state is observed. The primary resistive transition is observed at 50K, but the derivative of the resistivity data (inset, fig 6) shows that the transition observed in the susceptibility near 88K also impacts the resistivity. The behavior of the resistivity in the three-layer phase is very similar to that of two-layer $Na_{0.5}CoO_2$, though in the latter case the resistive transition often shows an initial increase near 50K with the main transition at lower temperatures.

**Heat Capacity.** The heat capacity data as a function of x for the three-layer derived $Na_xCoO_2$ phases are presented in figure 7. The specific heat for all compositions decreases as the temperature approaches 2K. For x=0.75, a small transition is seen around 25K, associated with the magnetic ordering observed in the $\chi$ vs T data. The electronic contribution to the specific heat ($\gamma$) was extracted from the Debye formula at temperatures lower than 10K. The inset of the figure shows $\gamma$ for each composition. As expected, the lowest $\gamma$ is found for x=0.5 where a transition from metallic to insulator takes place. Below and above this composition, the larger carrier concentrations are expected to result in larger values for $\gamma$.

**Discussion**

In contrast to the relatively straightforward structural behavior of the frequently studied two layer $Na_xCoO_2$ system, significant structural changes take place when the sodium composition is varied in three-layer $Na_xCoO_2$. In the two layer system, P2 type phases exist over a large composition range, dominating the phase diagram from x=0.3 to x=1. In that two layer system, x=0.5 is a special composition structurally, and there are small two-phase regions between different P2 type structures in the high sodium content region. The Na is in trigonal prismatic coordination across the whole two-layer series, resulting in the fact that neighboring $CoO_2$ layers are stacked in the same position relative to each other in all compounds. The same is not the case for the three-layer family. Our studies show that octahedral coordination for Na is apparently destabilized near half filling (i.e. x ~ 0.5) of the Na planes. The energetic reason for this is not known. The appearance of P1 phases in the middle of the three-layer phase diagram due to this octahedral site destabilization is the major factor in complicating the structural phase diagram. Thus, unlike the two-layer system, the regions of solid solution in the three-layer $Na_xCoO_2$ are relatively narrow and compositions intermediate to those described here are classic two-phase mixtures.

The changes in Na coordination across the series cause the $CoO_2$ layers to shift relative to each other to create the appropriate shape coordination polyhedra for Na and the right position of the adjacent O layers. The Na coordination polyhedra observed are presented in figure 8. Both the bond lengths and shapes of the octahedra and trigonal prisms are consistent with expectations for Na - O polyhedra and very similar to what is seen in the two-layer series.[7] As in the two-layer series, the $NaO_2$ plane layer expands as Na is removed: when the sodium site occupancy within the layers decreases, the repulsion between the $CoO_2$ sheets is enhanced, and also, the coulombic forces holding the layers together decrease. Figure 9 shows that in spite of the changes in coordination of the Na across the series, the thickness of the $NaO_2$ layers changes continuously with Na content.

A characterization of the relation of the $CoO_2$ planes relative to one another across the series is presented in figure 10. In the upper panel, the distances between the $CoO_2$ planes (from Co plane to Co plane) are shown for all compounds. There is a uniform change to smaller separation with increasing Na concentration, due primarily to the changing concentration of Na in the $NaO_2$ plane, as described above.

The structures change symmetry across the series and the Na coordination changes from octahedral to trigonal prismatic and back again. A uniform description of the family can be made, however, by defining a pseudomonoclinic cell for all cases, with $a= a_{hex}$ and $b=a_{hex}\sqrt{3}$ for the in-plane cell parameters, and $c$ the distance from one Co to its equivalent Co one layer away. A pseudomonoclinic β angle gives the angle between the $c$ and $a$ pseudomonoclinic axes, and is a good measure of the relative positions of neighboring layers. As an example, figure 11 shows the pseudomonoclinic cell derived from the hexagonal cell for x=0.92 and x=0.75.

To characterize how the planes shift relative to one another across the series, the pseudomonoclinc angle and the actual distance of the plane shift on going from one compound to the next are shown in the bottom two panels of figure 10. The middle panel suggests that the x=0.75 compound has an unexpectedly large shift of the layers relative to the behavior of the rest of the series, where the angles change continuously with Na concentration. This unusual character for the x=0.75 phase is also suggested by the bottom panel: though the coordination of the Na remains octahedral when decreasing the Na content from x=0.92 to x=0.75, the layer has shifted by 0.6 Å. Interestingly, this shift is of the same magnitude, though in the opposite sense, as occurs when the layers shift between x=0.75 and x=0.6 to accommodate a change in the Na coordination from octahedral to prismatic. A somewhat smaller shift is observed on going from the trigonal prismatic to octahedral coordination between x=0.51 and 0.32. It is interesting that in all cases the magnitudes of the shifts are relatively small, 0.6 Å in the largest case.

As sodium content is varied in $Na_xCoO_2$, the formal oxidation state of Co must change. For ideal $NaCoO_2$, $Na_{0.5}CoO_2$ and $CoO_2$, for example, the Co formal oxidation states will be 3+, 3.5+, and 4+, respectively. Thus, like in the copper oxide superconductors where the charge in the $CuO_2$ planes can be deduced from the oxygen content or electropositive non transition metal ratios (e.g. in $La_{2-x}Ba_xCuO_4$), the sodium content can be used as a measure of the charge state of the electronic system in the $CoO_2$ planes in $Na_xCoO_2$. There are two caveats: the strict use of Na content for this purpose in two layer $Na_xCoO_2$ has been called into question for compositions where x is less than 0.5 by titration measurements,[15] and, secondly, like in other highly oxidized transition metal compounds, the formal charge value says nothing about how the excess positive charge, on going from $Co^{3+}$ to $Co^{4+}$, is distributed among the Co or its coordinating oxygens. Plotting of structural parameters relevant to the electronic system as a function of Na content is straightforward, however, and can be used to infer general characteristics of the electronic system.

The structural characterization of the $CoO_2$ plane as a function of Na content in three-layer $Na_xCoO_2$ is presented in figure 12. As the sodium composition decreases, the in-plane Co-Co distance decreases, as seen in the upper panel. In the simplest picture, this is a result of increasing the formal oxidation state of cobalt from mainly $Co^{3+}$ in $Na_{0.92}CoO_2$ to $Co^{+3.68}$ in $Na_{0.32}CoO_2$: the in-plane size of the $CoO_6$ octahedra is expected to shrink as the $Co^{4+}/Co^{3+}$ ratio increases. It is of interest that the size change, though monotonic, is not linear across the series, but changes most in the mid composition regions. Structural study of the two-layer $Na_xCoO_2$ system has lead to the suggestion that the thickness of the $CoO_2$ layers may be a good reflection of redistribution of charge among different electronic orbitals with variation of electron count in that system.[7] The thickness of the $CoO_2$ layer as a function of sodium content in the three-layer system is shown in the middle panel of figure 12. As expected, for x=0.92 where the cobalt ions are mainly $Co^{3+}$, the $CoO_2$ layer thickness is largest, because the octahedra are largest, while for x=0.32, where the formal oxidation state is $Co^{3.68+}$, the $CoO_2$ layer thickness is smallest. The figure shows (middle panel), however, that the variation in layer thickness, in this series is not monotonic, suggesting that there is a significant redistribution of charge among different electronic orbitals across the three-layer series as a function of Na content. Significantly, the Co-O bond length varies continuously across the series, reflecting a systematic, continuous change in the formal Co oxidation state. The data presented in this figure suggest that in three-layer $Na_xCoO_2$, the electronic system changes, and the crystal structure responds, over the whole composition region, from x=0.32 to x=0.92.

Finally, figures 13 and 14 present a general comparison of the electronic characteristics of the $CoO_2$ planes as a function of Na concentration in the two-layer and three-layer systems. Figure 13 shows that the variation in Co-O bond length across the two series, are very similar, and follow the expected trend toward larger size with decreased Co oxidation state. There do appear to be subtle differences observed, but more detailed study would be required to clarify them. This type of continuous bond length change with x would not be expected if oxygen vacancies occurred in significant numbers for $x<0.5$, halting the possible oxidation state of Co at an upper limit of 3.5+.

Figure 14 shows the thickness of the $CoO_2$ layer relative to that expected if the $CoO_6$ octahedra have an ideal shape, as a function of composition. This quantity is determined from the length of the edges of the in-plane face of the $CoO_6$ octahedra and the ideal geometric relationship between the edge length and the diagonal height of an octahedron. The $CoO_2$ layer is highly compressed from what is expected for ideal octahedra (only 85% of the ideal value) in both phase families, suggesting that the structural distortion should be large enough to strongly influence the relative energies of different Co 3d suborbitals. This thickness varies across the series, reflecting a redistribution of charge within the $CoO_2$ layers as a function of composition and comparison of the two-layer and three-layer phases suggests that there are subtle differences in the electronic systems. In particular, the figure suggests that the three-layer $Na_{0.75}CoO_2$ phase has a different kind of electronic structure than is seen in the two-layer variant at the same composition. Also shown in figure 14 are the single phase and multiple regions in the two families of compounds. This indicates how strongly the type of stacking influences the crystal chemistry of these systems. In addition, the figure illustrates a comparison of the Co positions in the two-layer and three-layer structure types. In the former, the Co planes are eclipsed while in the later the Co planes are staggered.

**Conclusions**

The crystal structures of the $Na_xCoO_2$ phases derived from three-layer $NaCoO_2$ are more complex than the much studied two-layer structures. This is due in large part to the fact that unlike the two-layer system, in the three-layer derived system the Na coordination changes in the structural series. In the two-layer family, the sodium forms many ordered phases, both commensurate and incommensurate with the underlying $CoO_2$ lattice. In the present system, due to the different types of sites encountered (i.e. octahedral rather than prismatic) the Na ordering may be expected to be different. No information on that ordering is provided in the average structure determinations presented here, and would be of interest in future studies. The structural complexity of the system is reflected in the crystal structure of the x=0.75 composition, where unlike the case of the analogous two-layer variant, the high intensities of the incommensurate superlattice reflections suggest that the underlying $CoO_2$ lattice experiences some kind of structural modulation. The determination of this structure will be of interest, as well as modeling to determine whether the structural modulation is electronically driven.

Magnetic susceptibility data in the current family are consistent with $Co^{3+}$ and $Co^{4+}$ ions being in low spin configuration, and the magnetic behavior is similar to that observed in the two-layer system, though some differences are seen; particularly the presence of a Curie-Weiss susceptibility in the x=0.3 sample. The three-layer derived x=0.5 phase, in its initial characterization reported here, appears to be analogous to the two-layer variant, suggesting that the electronic instability that gives rise to the insulating behavior is strongly two-dimensional in character, due to the fact that the stacking of the $CoO_2$ layers differs in the two systems. Finally, the composition dependence of the electronic contribution to the specific heat in the three-layer system appears to be substantially different from what has been reported in the two-layer system;[16, 17,18] in the present case showing the largest $\gamma$ value at x=0.6. This may be due to the special character of the x=0.75 composition in the three-layer system, which may suppress $\gamma$. These similarities and differences suggest that further work on the three-layer derived phases and comparison to the two-layer phases will be of interest.

**Acknowledgements**

The work at Princeton was supported by the Department of Energy, grant DOE-FG98-0244254, and by the National Science Foundation, grant DMR – 0213706.

**Table 1. The Cell Parameters for $Na_xCoO_2$ (x=0.92, 0.6, 0.48 and 0.32).**

| Compound | Space Group | Cell Constants (Å) | Volume ($Å^3$) | Volume/F.U. ($Å^3$) |
|---|---|---|---|---|
| $Na_{0.92}CoO_2$ | *R -3 m* (No 166) | *a* = 2.88878(5)<br>*c* = 15.5998(3) | 112.740(5) | 37.58 |
| $Na_{0.75}CoO_2$ | *C 2/m* (No.12) | *a* = 4.9020(5)<br>*b* = 2.8723(3)<br>*c* = 5.7789(7)<br>*β* = 111.764(7) | 75.57(1) | 37.79 |
| $Na_{0.60}CoO_2$ | *C2/ m* (No 12) | *a* = 4.9043(2)<br>*b* = 2.8275(1)<br>*c* = 5.7097(3)<br>*β* = 106.052(3) | 76.089(6) | 38.05 |
| $Na_{0.51}CoO_2$ | *C2/ m* (No 12) | *a* = 4.8809(1)<br>*b* = 2.81535(9)<br>*c* = 5.7738(2)<br>*β* = 105.546(2) | 76.439(4) | 38.22 |
| $Na_{0.32}CoO_2$ | *R -3 m* (No 166) | *a* = 2.81202(7)<br>*c* = 16.732(1) | 114.58(1) | 38.18 |

**Table 2. Crystallographic data for $Na_{0.92}CoO_2$ in the space group *R -3 m* (No 166)**

| Atom | Wyckoff position | *x* | *y* | *z* | *Uiso**100 | *Occ* |
|---|---|---|---|---|---|---|
| Co | 3a | 0 | 0 | 0 | 0.43(3) | 1 |
| Na | 3b | 0 | 0 | 0.5 | 0.72(4) | 0.921(7) |
| O | 6c | 0 | 0 | 0.26976(3) | 0.91(1) | 1 |

$\chi^2$=1.39, $wR_p$=4.95%; $R_p$=4.26%

**Table 3. Crystallographic data for $Na_{0.32}CoO_2$ in the space group *R -3 m* (No 166)**

| Atom | Wyckoff position | *x* | *y* | *z* | *Uiso**100 | *Occ* |
|---|---|---|---|---|---|---|
| Co | *3a* | 0 | 0 | 0 | 0.78(6) | 1 |
| Na | *3b* | 0 | 0 | 0.5 | 1.2(2) | 0.32(1) |
| O | *6c* | 0 | 0 | 0.2762(1) | 1.07(4) | 1 |

$\chi^2$=1.19, $wR_p$=5.26%; $R_p$=4.33%

**Table 4. Crystallographic data for $Na_{0.5}CoO_2$ in the space group *C2/m* (No 12)**

| Atom | Wyckoff position | *x* | *y* | *z* | *Uiso**100 | *Occ* |
|---|---|---|---|---|---|---|
| Co | *2a* | 0 | 0 | 0 | 0.86(5) | 1 |
| Na | *4i* | 0.806(2) | 0 | 0.491(2) | 2.1(2) | 0.256(7) |
| O | *4i* | 0.3871(3) | 0 | 0.1740(4) | 0.92(3) | 1 |

$\chi^2$=1.48, $wR_p$=4.51%; $R_p$=3.74%

**Table 5: Crystallographic data for $Na_{0.6}CoO_2$ in the space group *C 2/m* (No 12)**

| Atom | Wyckoff position | *x* | *y* | *z* | *Uiso**100 | *Occ* |
|---|---|---|---|---|---|---|
| Co(1) | *2a* | 0 | 0 | 0 | 1.01(5) | 1 |
| Na(1) | *8j* | 0.812(2) | 0.049(6) | 0.493(1) | 1.6(3) | 0.149(3) |
| O(1) | *4i* | 0.3886(2 | 0 | 0.1792(3) | 0.94(2) | 1 |

$\chi^2$=2.25, $wR_p$=5.84%; $R_p$=4.72%

**Table 6: Crystallographic data for $Na_{0.75}CoO_2$ in the space group *C 2/m* (No 12)**

| Atom | Wyckoff position | *x* | *y* | *z* | *Uiso**100 | *Occ* |
|---|---|---|---|---|---|---|
| Co(1) | *2a* | 0 | 0 | 0 | 4.0(2) | 1 |
| Na(1) | *2d* | 0 | 0.5 | 0.5 | 5.3(3) | 0.75 |
| O(1) | *4i* | 0.2576(7) | 0 | 0.8179(6) | 2.16(8) | 1 |

**Table 7. Summary of Magnetic Data for three-layer $Na_xCoO_2$.**

| $Na_xCoO_2$ | C ($cm^3K$/mole) | $\mu_{eff}$ ($\mu_B$) | θ (K) | $\chi_0$ (emu/$mole_{Co}$Oe) |
|---|---|---|---|---|
| x=0.92 | 0.0130(5) | 0.32 | -55(±3) | 0.00008(1) |
| x=0.75 | 0.0984(9) | 0.88 | -48(±1) | 0.00011(1) |
| x=0.6 | 0.084(3) | 0.82 | -97±3) | 0.00037(8) |

**References**

[1]M.G.S.R. Thomas, P.G. Bruce, J.B. Goodenough, *Solid State Ionics*, **17**, 13, (1985).

[2]K. Takada, H. Sakurai, E. Takayama-Muromachi, F. Izumi, R.A. Dilanian, and T. Sasaki, *Nature*, **422**, 53 (2003)

[3]I. Terasaki, Y.Sasago, K.Uchinokura, *Phys.Rev. B***56,** 12685, (1997)

[4]Y.Wang, N.S.Rogado, R.J.Cava and N.P.Ong, *Nature,* **423**, 425 (2003)

[5]M.L.Foo, Y.Wang, S.Watauchi, H.W. Zandbergen, T. He, R.J. Cava, N.P.Ong, *Phys. Rev. Lett.* **92**, 247001 (2004).

[6]Claude Fouassier, Guy Matejka, Jean-Maurice Reau, Paul Hagenmuller, *J. Solid State Chem.* **6**, 532 (1973).

[7]Q. Huang, M.L.Foo, R.A. Pascal, Jr., J.W.Lynn, B.H.Toby, T. He, H.W. Zandbergen, R.J.Cava, *Phys. Rev B* **70,** 184110 (2004).

[8]Y. Takahashi, Y.Gotoh, J. Akimoto, *J. Solid State Chem.* **172**, 22 (2003).

[9]Y. Ono, R. Ishikawa, Y. Miyazaki, Y. Ishii, Y. Morii, T. Kajitani, *J. Solid State Chem.* **166,** 177 (2002)

[10]S. Kikkawa, S. Miyazaki, M. Koizumi, *J. Solid State Chem*, **62**, 35, (1986).

[11]Larson, A.; Von Dreele, R. B. GSAS: Generalized Structure Analysis System; Los Alamos National Laboratory: Los Alamos, NM (1994)

[12]Certain commercial product or equipment, is identified in this report to describe the subject adequately. Such identification does not imply recommendation or endorsement by the NIST, nor does it imply that the equipment identified is necessarily the best available for the purpose.

[13]C. Delmas, J.J. Braconnier, C. Fouassier, P. Hagenmuller, *Solid State Ionics*, **3/4**, 165 (1981).

[14]J. L. Gavilano, D. Rao, B. Pedrini, J. Hinderer, H.R. Ott, S.M. Kazakov, J. Karpinski, *Phys. Rev. B* **69**, 100404 (2004)

[15]M. Karppinen, I. Asako, T. Motohashi, H. Yamauchi, *Phys. Rev. B* **71**, 092105 (2005)

[16]M. Yokoi, T. Moyoshi, Y. Kobayashi, M. Soda, Y. Yasui, M. Sato, K. Kakumai, *Cond-mat*/0506220

[17]R. Jin, B.C. Sales, S.Li and D. Mandrus, *Phys Rev B* **72**, 060512(R) (2005).

[18]M. Brühwiler, B. Batlogg, S. M. Kazakov, J. Karpinski, *cond-mat*/0309311

**Figure Captions**

**Figure 1.** The crystal structures of $Na_xCoO_2$ phases (x=0.92, 0.75, 0.6, 0.48, 0.32) derived from three-layer $NaCoO_2$. Smaller and bigger black spheres represent Co and sodium ions respectively while the grey spheres are the oxygen ions.

**Figure 2.** Observed (crosses) and calculated (solid line) neutron diffraction intensities for single-layer $Na_{0.51}CoO_2$ at 295K. Vertical bars show the Bragg peak positions. The difference plot is shown at the bottom.

**Figure 3**. Observed (crosses) and calculated (solid line) X-ray diffraction intensities for $Na_{0.75}CoO_2$. Vertical bars show the Bragg peak positions. The inset shows the superreflection peaks (marked with arrows) in the neutron diffraction pattern indexed in a 4D cell.

**Figure 4.** [010] diffraction pattern of $Na_{0.75}CoO_2$. Strong superreflections are present. The diffraction pattern was taken with the beam partly on a relatively thick area and partly over the adjacent hole, which configuration resulted in a tail of the reflections to the upper left corner. The indexing of some of the reflections is given. The vector **q**, describing the satellites, is 0.330**a*** -0.247**c***.

**Figure 5.** Temperature dependence of the magnetic susceptibility for $Na_xCoO_2$ with x=0.92, 0.75, 0.6, 0.51, and 0.3.

**Figure 6.** The temperature dependence of the resistivity in the plane parallel to the $CoO_2$ layers in a single crystal of x=0.5. The inset shows the derivative curve.

**Figure 7.** Temperature dependence of the specific heat for $Na_xCoO_2$ with x=0.92, 0.75, 0.6, 0.51, and 0.32.

**Figure 8.** The $NaO_6$ coordination polyhedra found in the $Na_xCoO_2$ structures.

**Figure 9.** Top: The thickness of the $NaO_2$ layer as a function of x.

**Figure 10.** Top: The distance between neighboring $CoO_2$ layers (from Co plane to Co plane); Middle: Pseudo-monoclinic cell angle; and Bottom: The layer shift in $Na_xCoO_2$ phases; as a function of x in three-layer $Na_xCoO_2$.

**Figure 11.** The pseudomonoclinic cell derived from the three-layer hexagonal structure for x=0.92 and x=0.75. The actual unit cell for both compositions is shown with thick lines.

**Figure 12.** Top: The in-plane Co-Co separation; Middle: the thickness of the $CoO_2$ layers; and Bottom: The variation of the Co-O bond length; as a function of x in three-layer $Na_xCoO_2$.

**Figure 13.** Comparison of the Co-O bond lengths as a function of x in $Na_xCoO_2$ in the two-layer and three-layer series. The line drawn is a guide to the eye.

**Figure 14.** Upper panel: the thickness of the $CoO_2$ plane relative to the thickness expected for an ideal $CoO_6$ octahedron (ideal thickness = 1); as a function of x in three-layer $Na_xCoO_2$. Lower panel: the same characteristic for the two-layer phase. Shaded regions are two phases regions. The nomenclature P2, O3, O1 and P1 are used to describe the Na ion coordination and the number of $CoO_2$ layers per cell (1, 2 or 3). H1, H2 and H3 refer to subtle differences in crystal structure within the P2 phase. The illustration on the top show a comparison of the Co position in the two-layer and three-layer variants. In the two layer phases the Co planes are the same in all layers. In the three-layer phases the triangular Co planes are staggered: black balls represent layer 1; dark grey balls represent layer 2 and light grey balls represent layer 3.

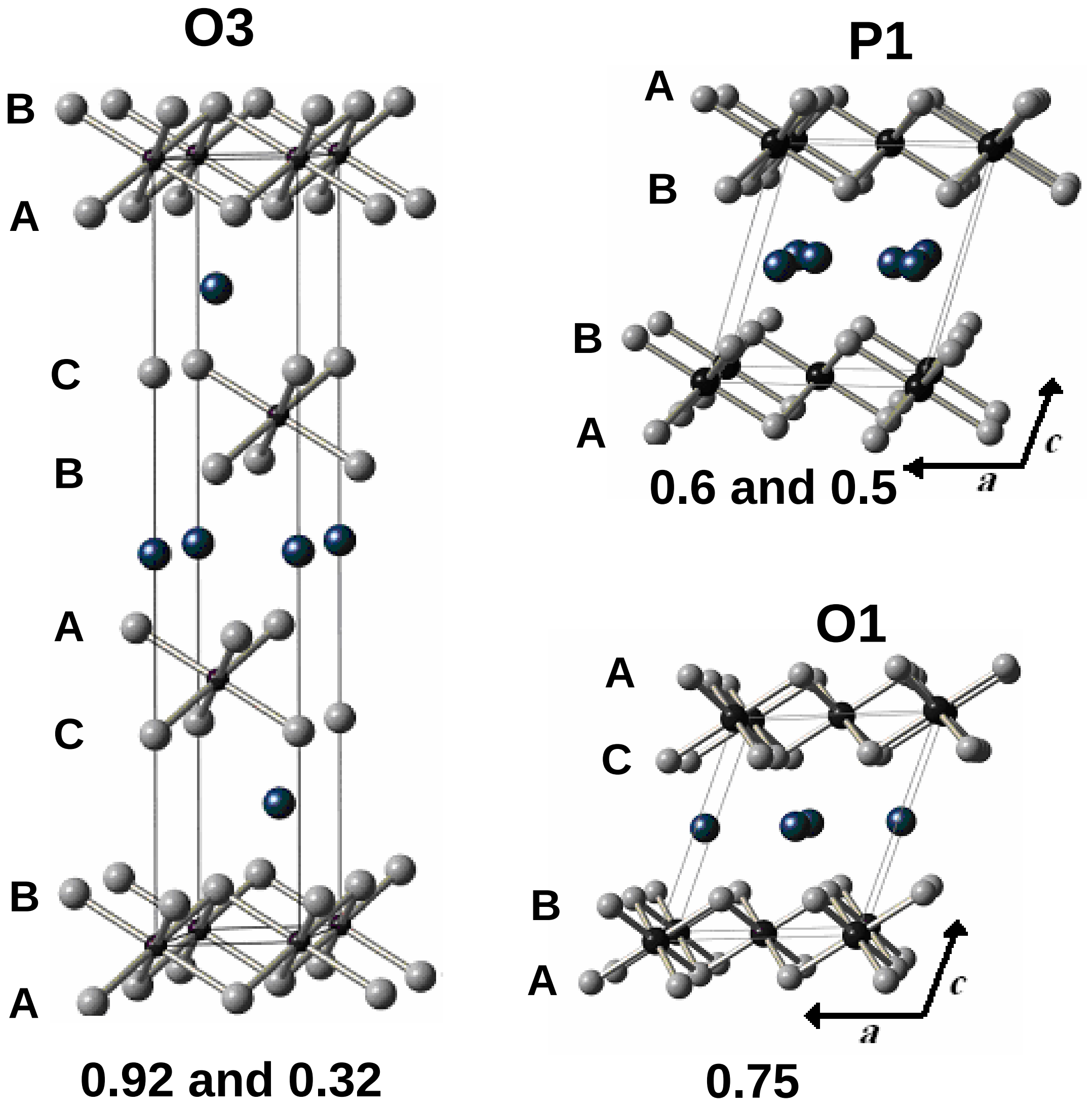

**Figure 1**

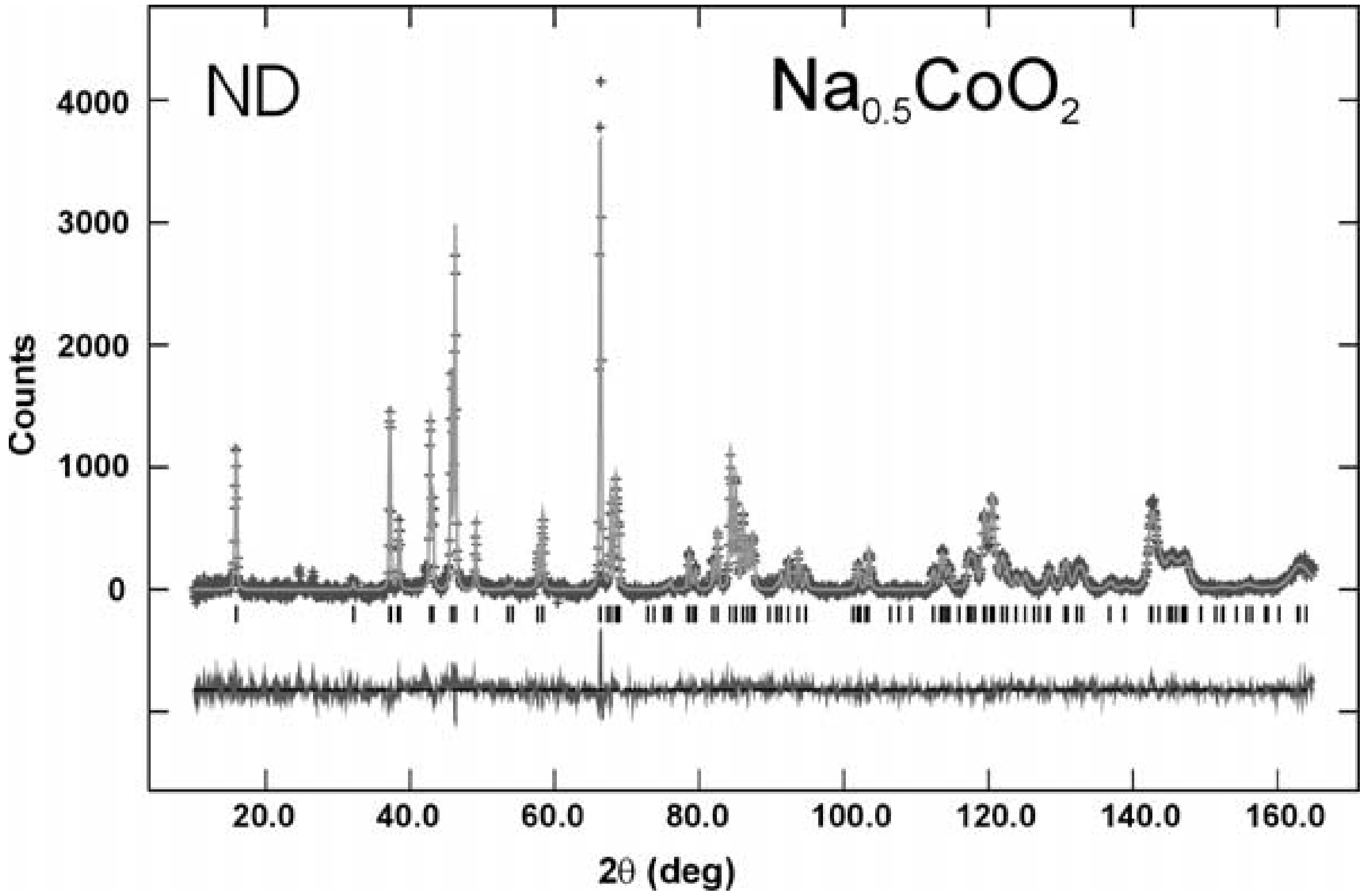

**Figure 2**

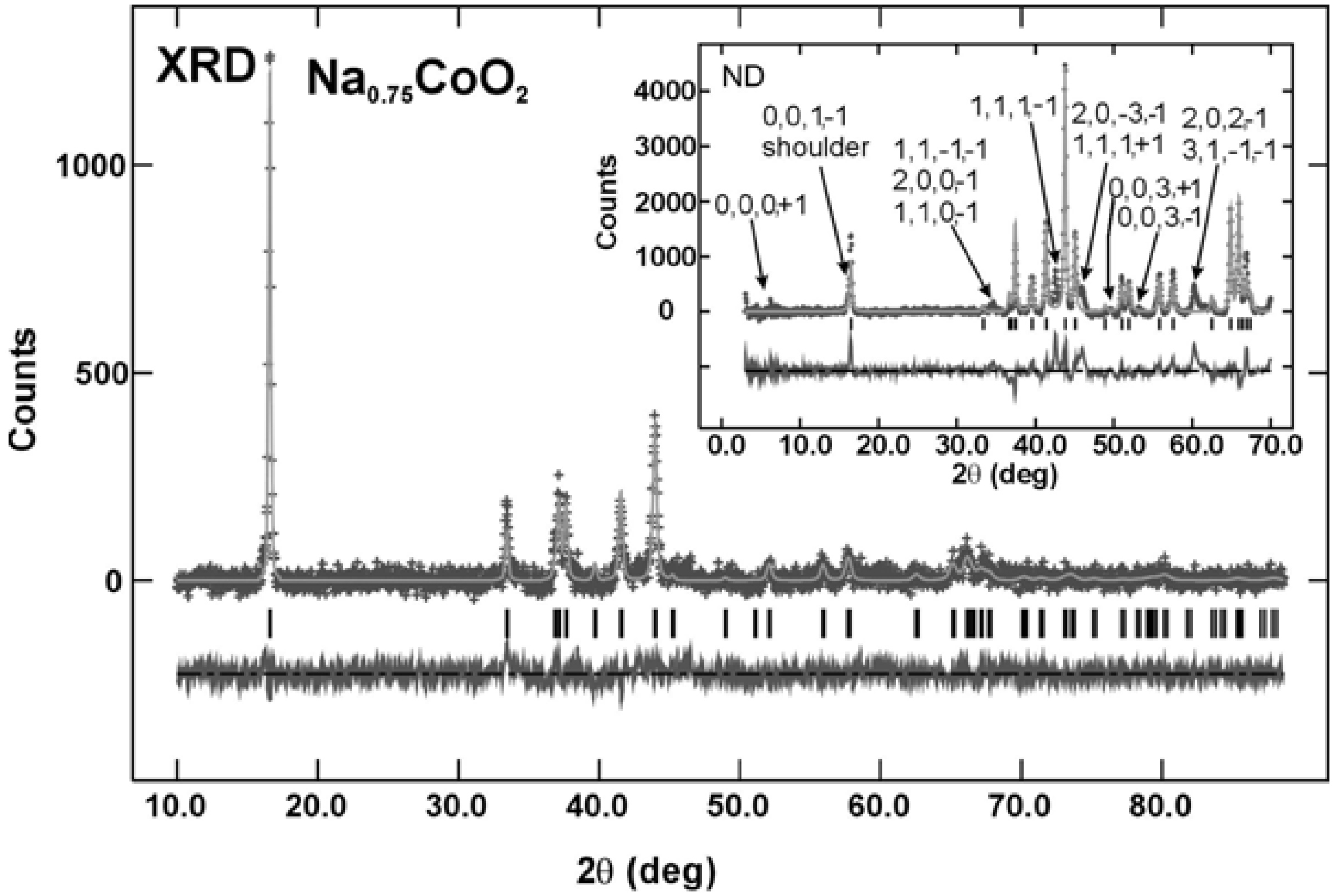

**Figure 3**

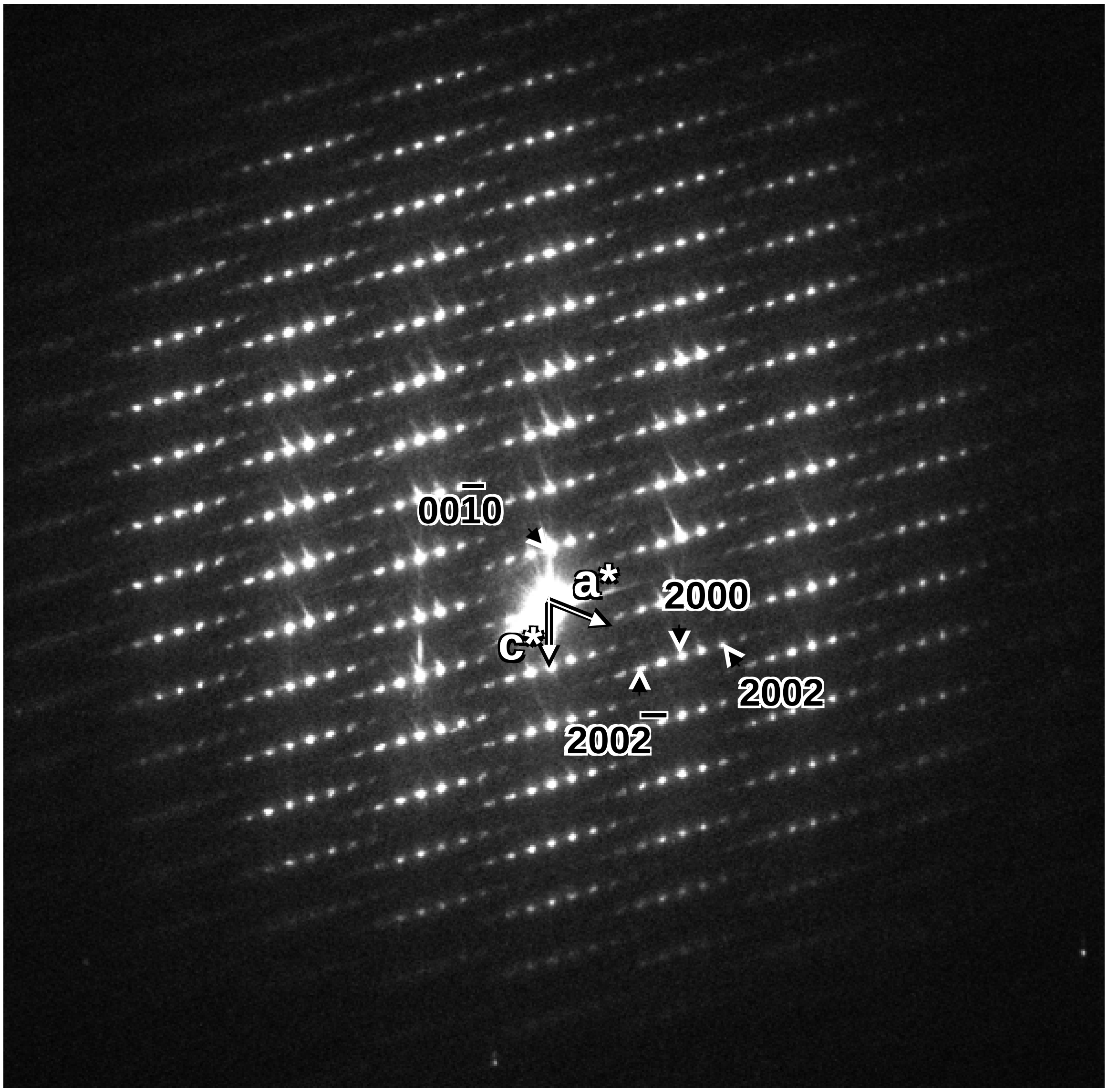

**Figure 4**

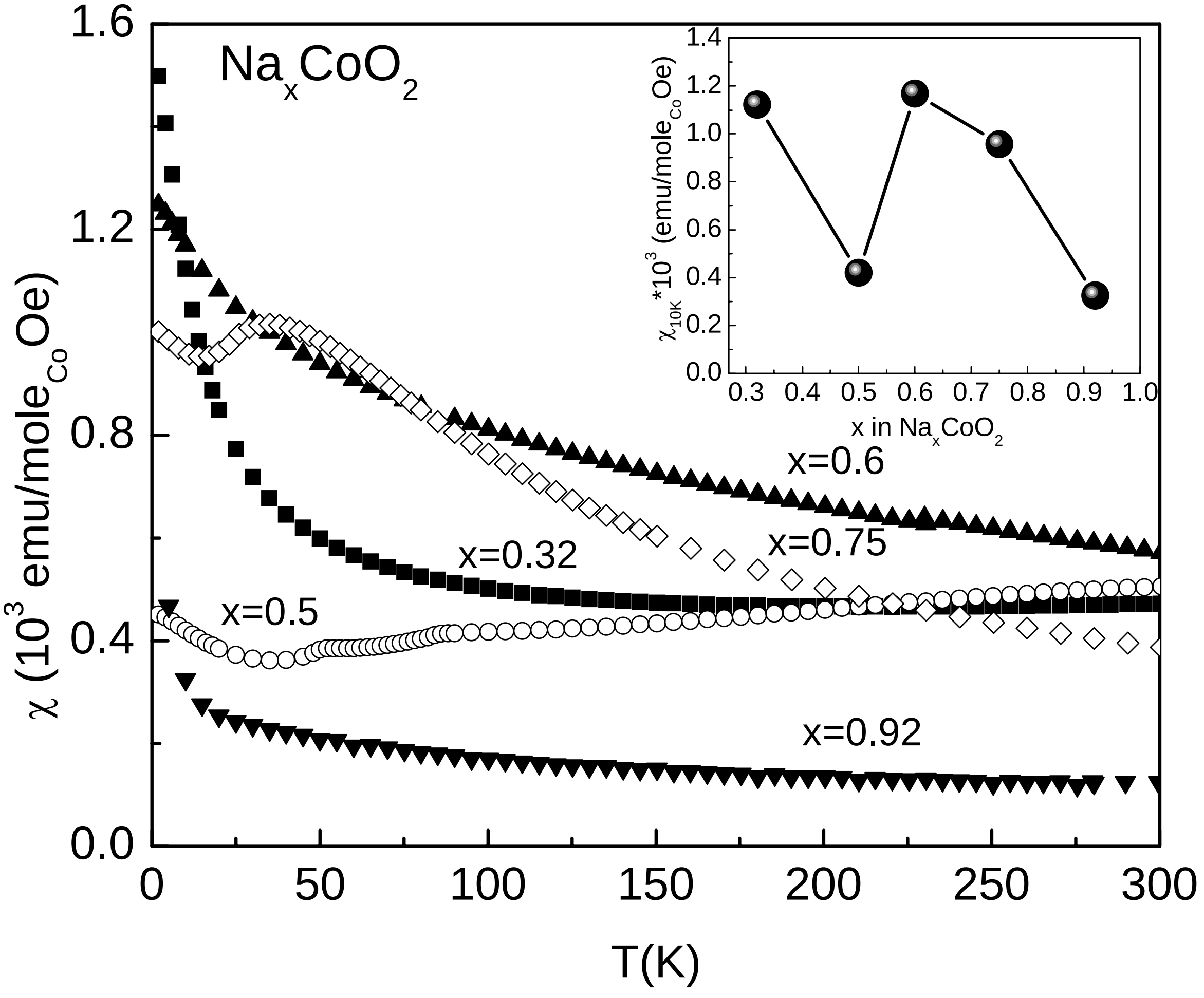

**Figure 5**

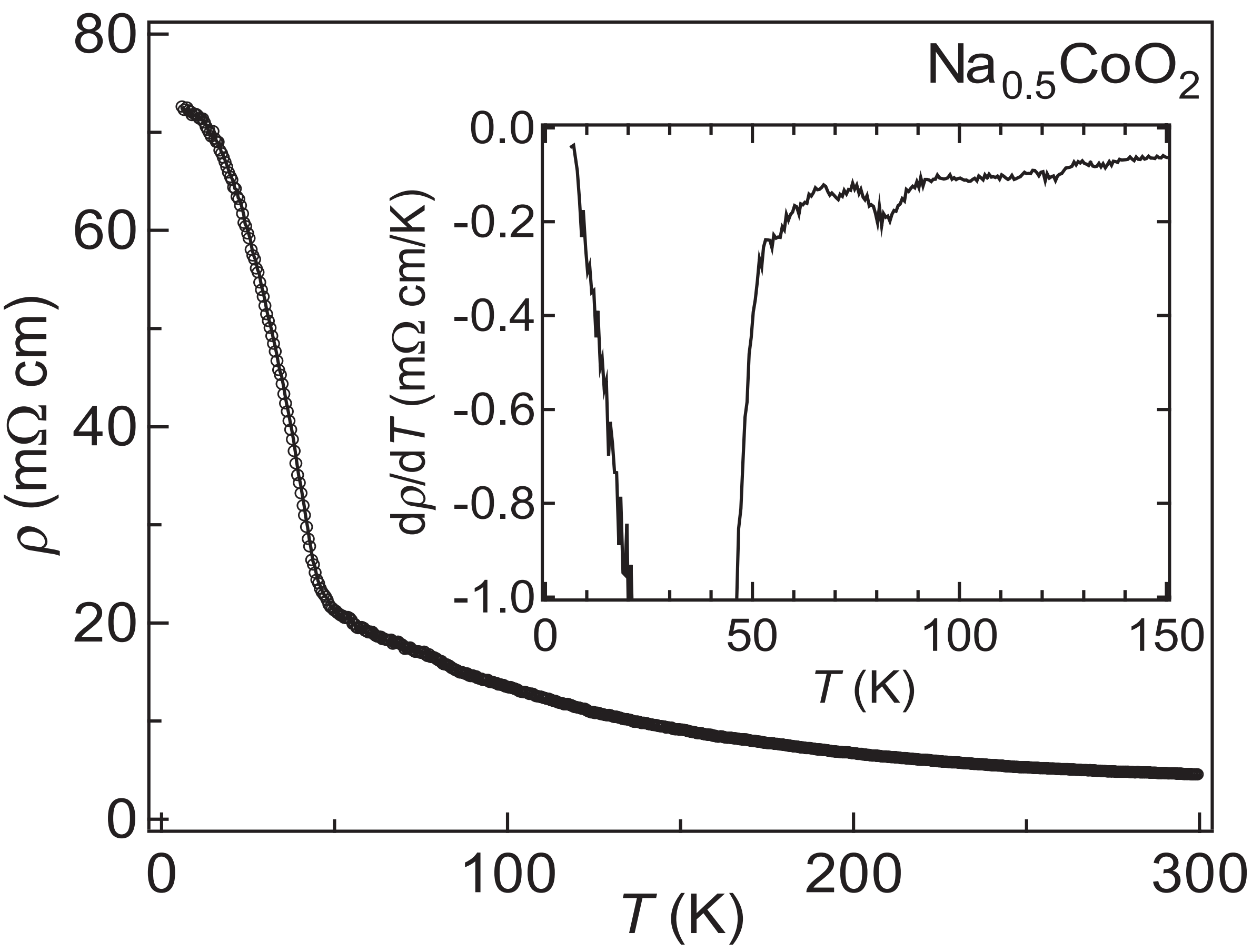

**Figure 6**

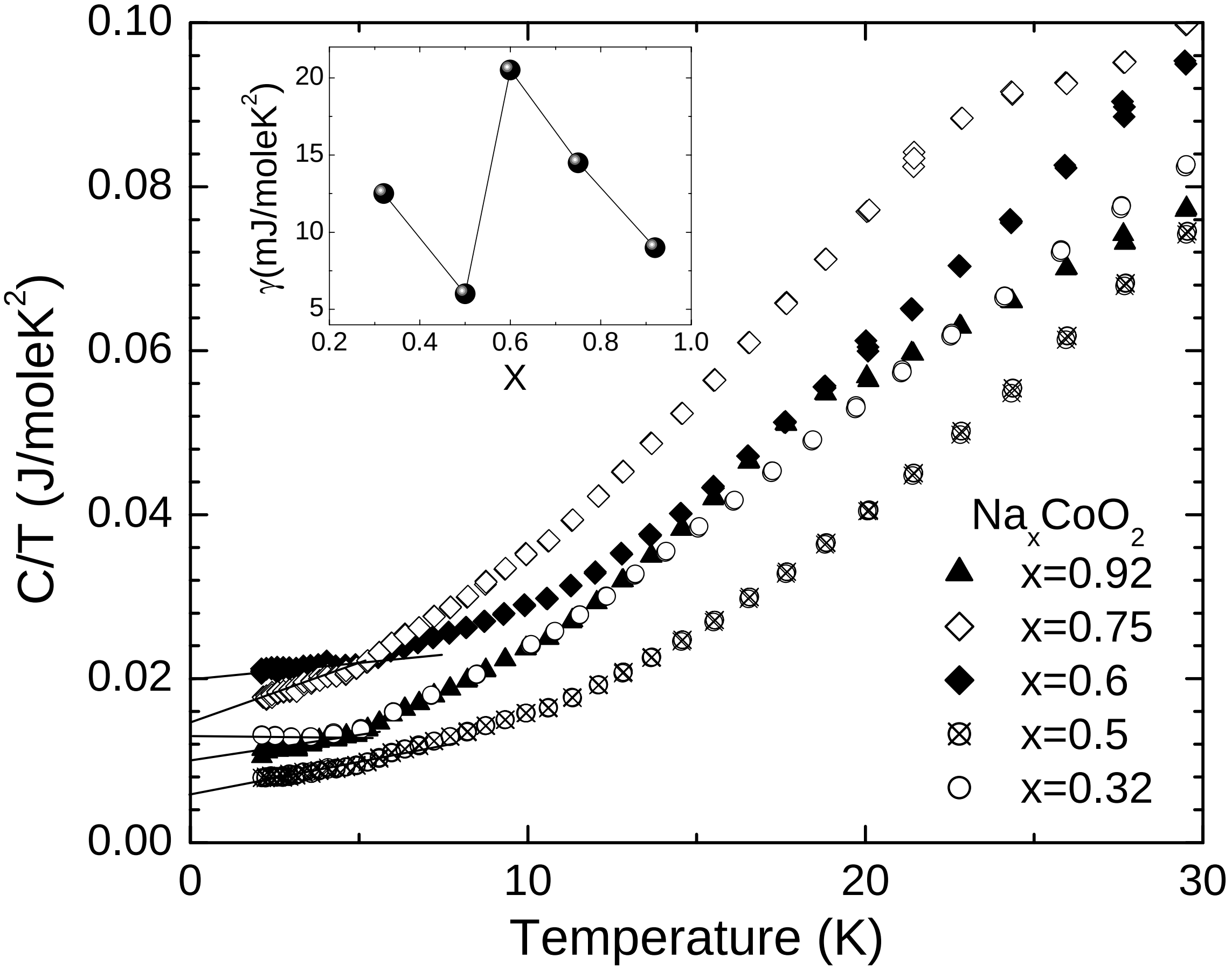

**Figure 7**

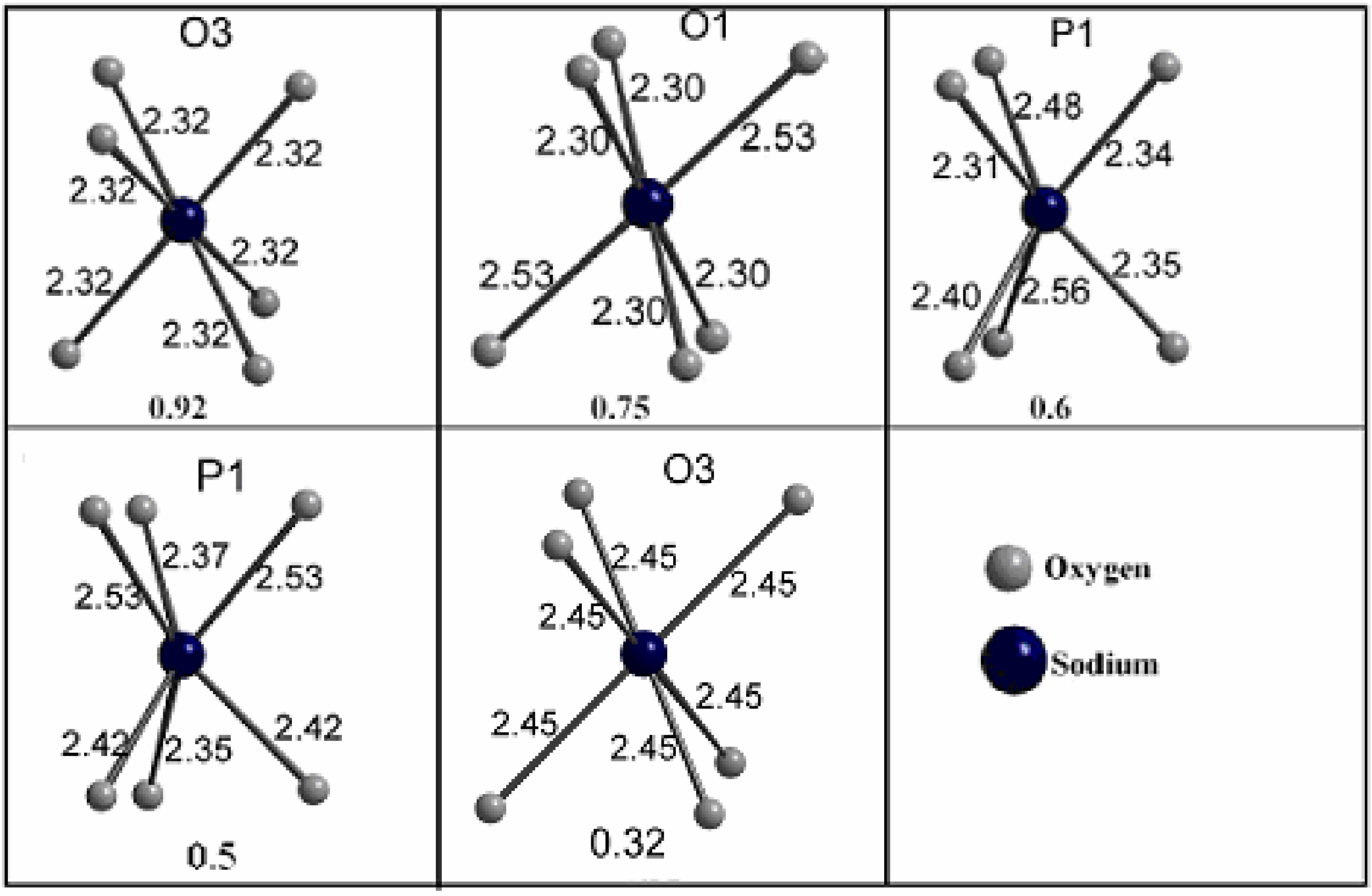

**Figure 8**

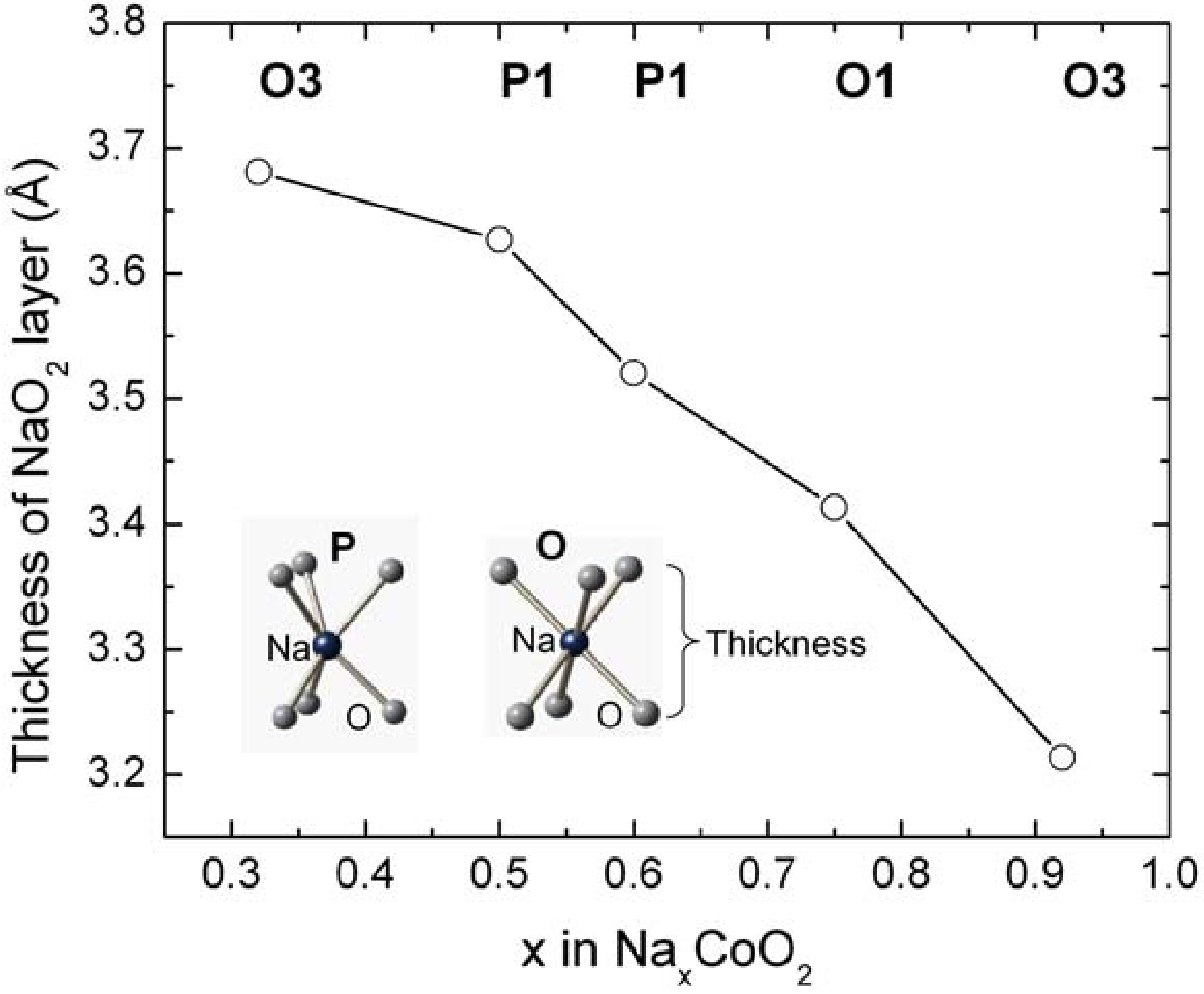

**Figure 9**

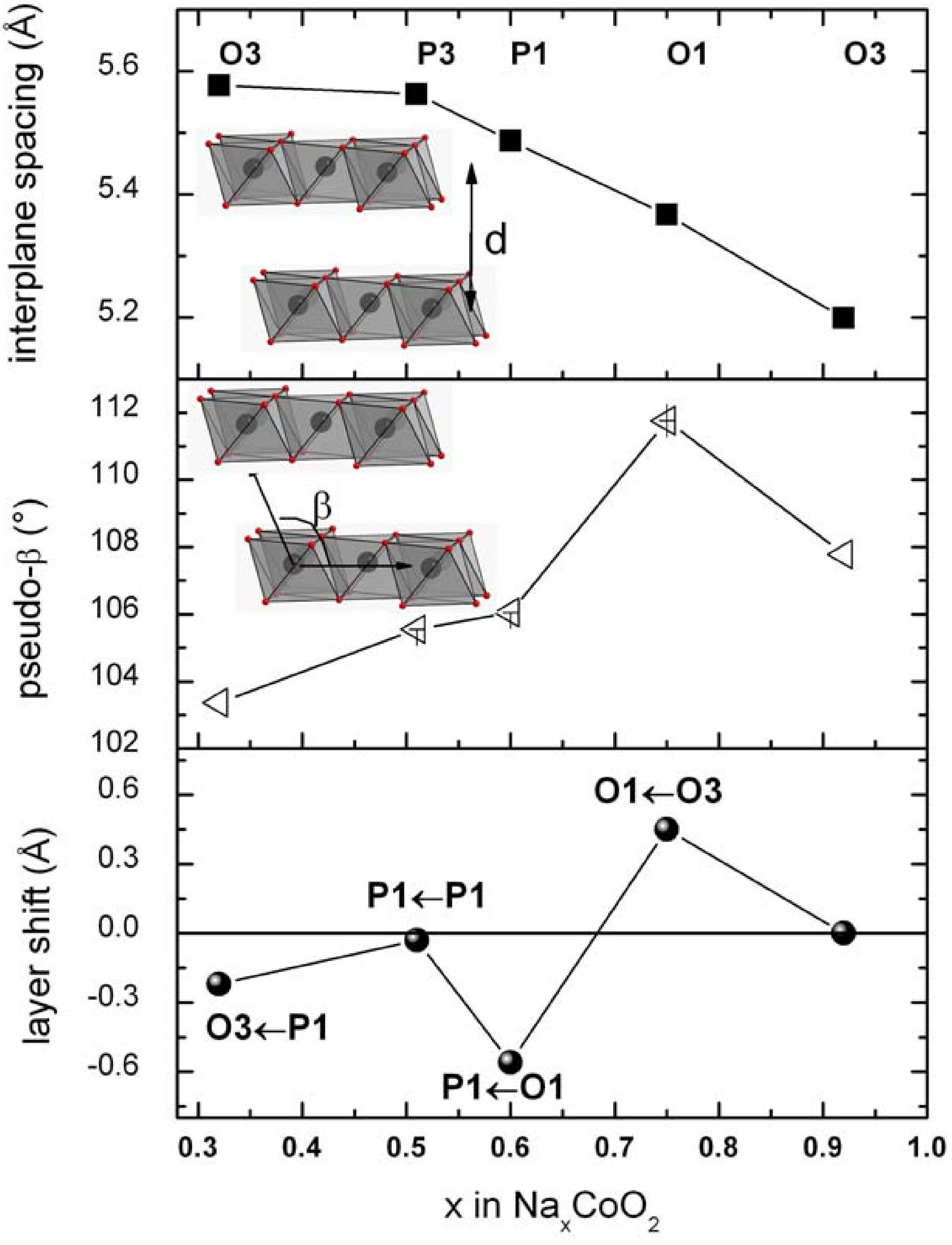

**Figure 10**

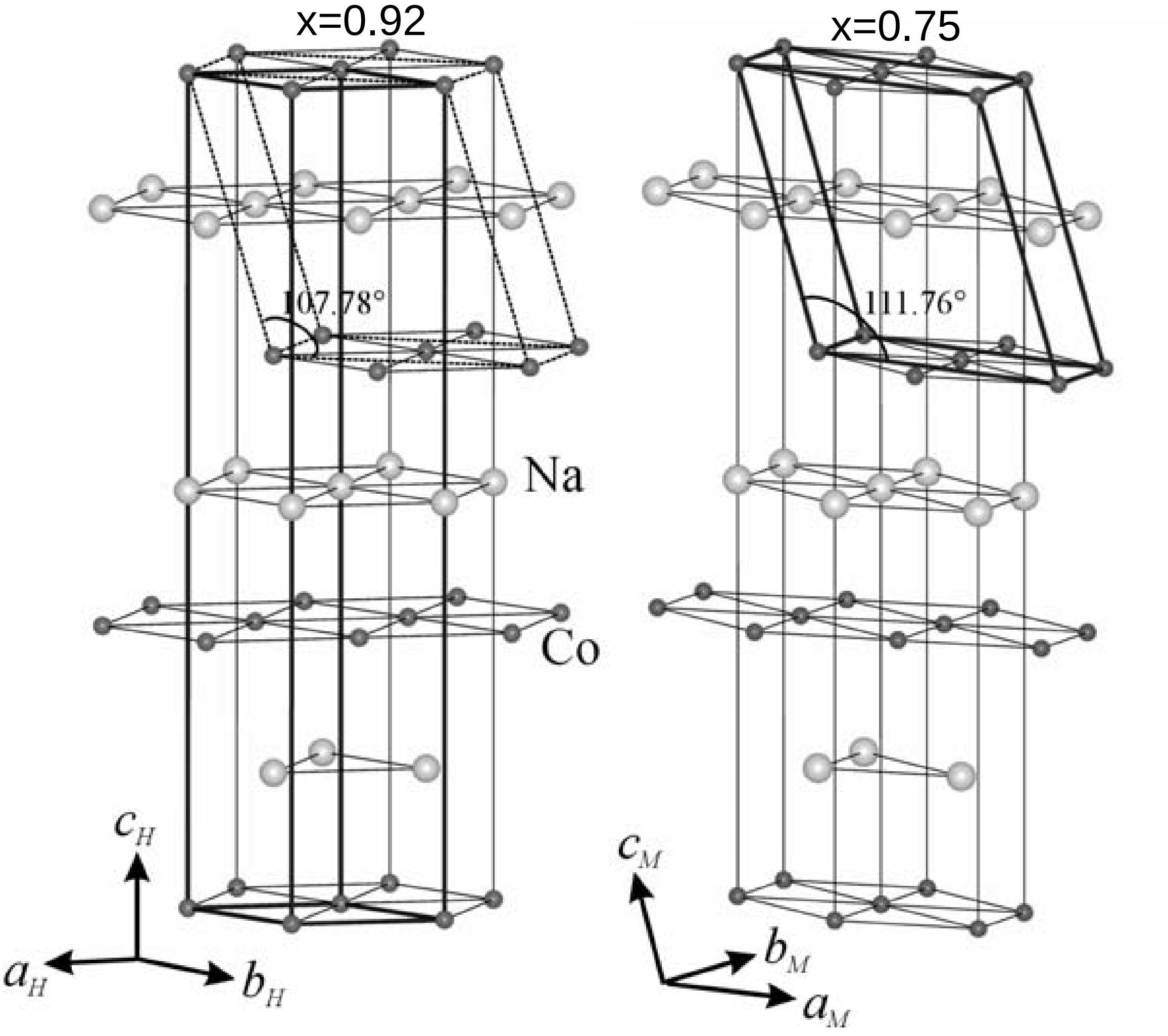

**Figure 11**

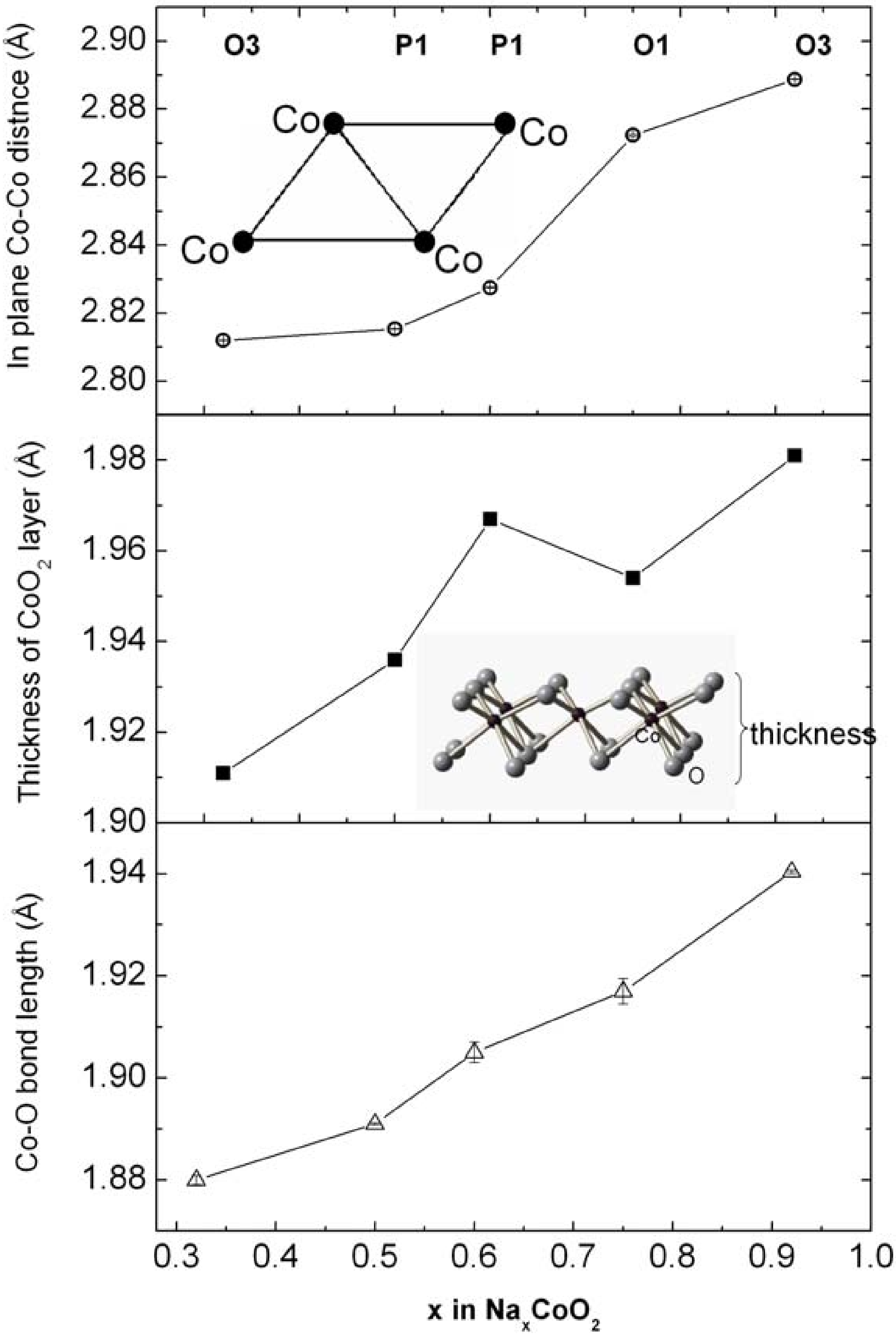

**Figure 12**

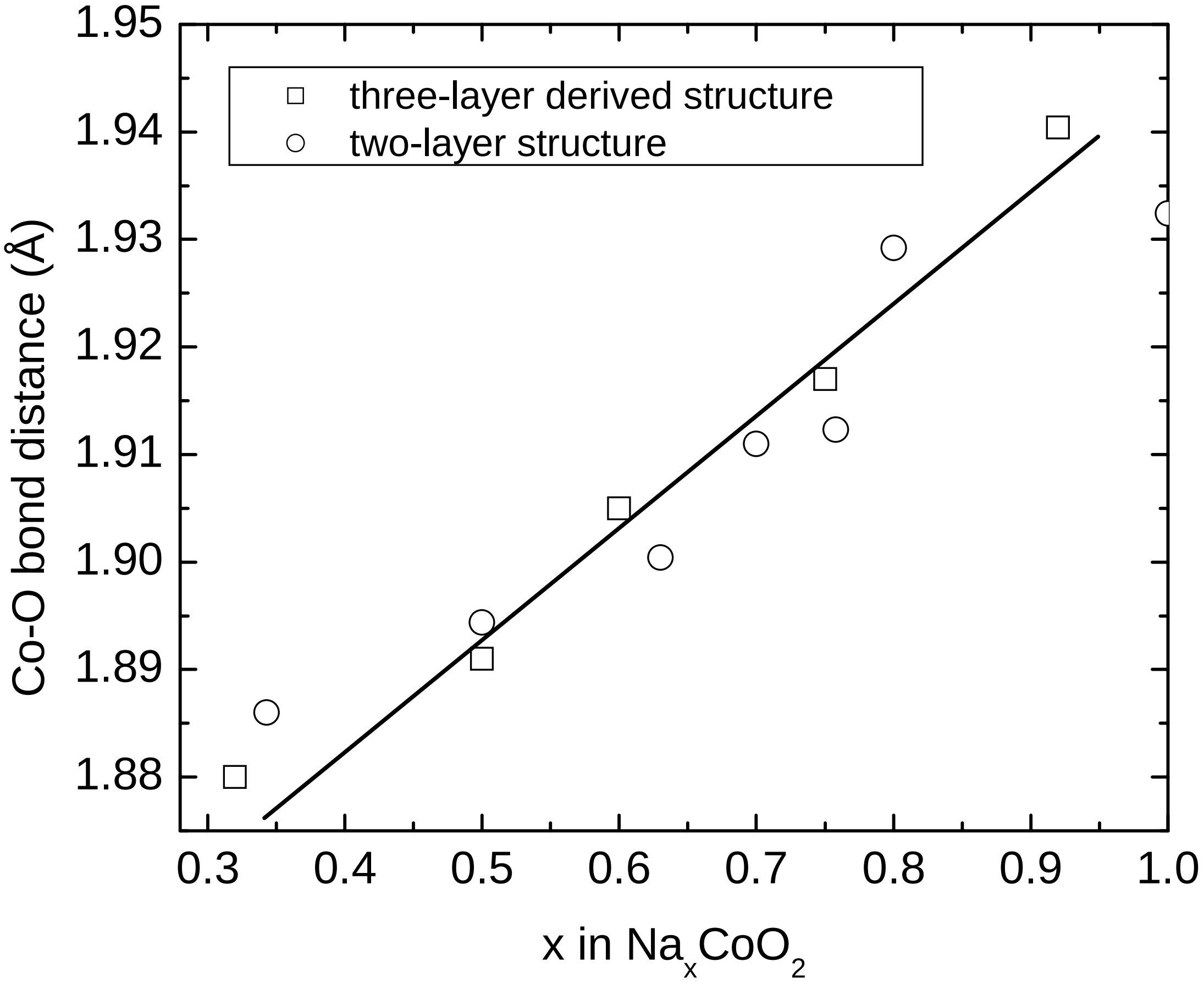

**Figure 13**

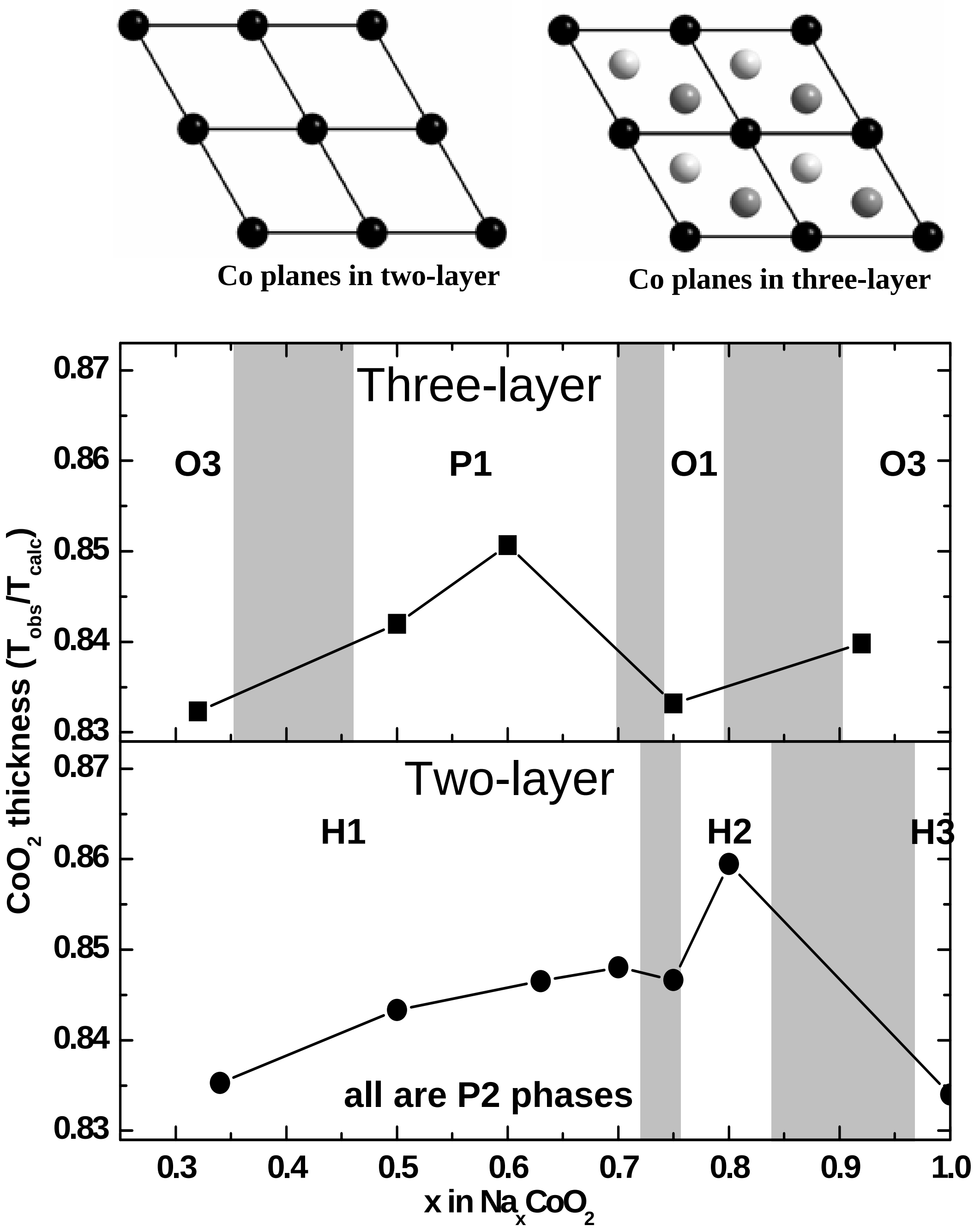

Figure 14